\makeatletter
\newcommand{\dontusepackage}[2][]{%
  \@namedef{ver@#2.sty}{9999/12/31}%
  \@namedef{opt@#2.sty}{#1}}
\makeatother
\dontusepackage{subfigure}

\documentclass[]{article}

\usepackage{lmodern}
\usepackage{amssymb,amsmath}
\usepackage{ifxetex,ifluatex}
\usepackage[usenames,dvipsnames]{color}
\usepackage{fixltx2e} 
\ifnum 0\ifxetex 1\fi\ifluatex 1\fi=0 
  \usepackage[T1]{fontenc}
  \usepackage[utf8]{inputenc}
\else 
  \ifxetex
    \usepackage{mathspec}
    \usepackage{xltxtra,xunicode}
  \else
    \usepackage{fontspec}
  \fi
  \defaultfontfeatures{Mapping=tex-text,Scale=MatchLowercase}
  
\fi
\IfFileExists{upquote.sty}{\usepackage{upquote}}{}
\IfFileExists{microtype.sty}{%
\usepackage{microtype}
\UseMicrotypeSet[protrusion]{basicmath} 
}{}
\usepackage[margin=1.0in,bottom=1.5in]{geometry}
\usepackage[square,numbers,sort&compress]{natbib}
\usepackage{listings}
\usepackage{graphicx}
\makeatletter
\def\maxwidth{\ifdim\Gin@nat@width>\linewidth\linewidth\else\Gin@nat@width\fi}
\def\maxheight{\ifdim\Gin@nat@height>\textheight\textheight\else\Gin@nat@height\fi}
\makeatother
\setkeys{Gin}{width=\maxwidth,height=\maxheight,keepaspectratio}
\usepackage{caption}
\usepackage{float}
\usepackage{subfig}
\ifxetex
  \usepackage[setpagesize=false, 
              unicode=false, 
              xetex]{hyperref}
\else
  \usepackage[unicode=true]{hyperref}
\fi
\hypersetup{breaklinks=true,
            bookmarks=true,
            pdfauthor={},
            pdftitle={Velocity continuation with Fourier neural operators for accelerated uncertainty quantification},
            colorlinks=true,
            citecolor=black,
            urlcolor=blue,
            linkcolor=black,
            pdfborder={0 0 0}}
\usepackage[preprint]{neurips_2019}
\usepackage{url}
\usepackage{booktabs}
\usepackage{amsfonts}
\usepackage{nicefrac}
\usepackage{microtype}

\newcommand{\B}{\mathbf}

\DeclareMathOperator*{\argmin}{arg\,min}

\title{Velocity continuation with Fourier neural operators for accelerated
uncertainty quantification}
\author{Ali Siahkoohi, Mathias Louboutin, and Felix J. Herrmann\\Georgia
Institute of
Technology\\\texttt{\{alisk,\phantom{\ }mlouboutin3,\phantom{\ }felix.herrmann\}@gatech.edu}}
\date{}

\begin{document}
\maketitle
\begin{abstract}
Seismic imaging is an ill-posed inverse problem that is challenged by
noisy data and modeling inaccuracies---due to errors in the background
squared-slowness model. Uncertainty quantification is essential for
determining how variability in the background models affects seismic
imaging. Due to the costs associated with the forward Born modeling
operator as well as the high dimensionality of seismic images,
quantification of uncertainty is computationally expensive. As such, the
main contribution of this work is a survey-specific Fourier neural
operator surrogate to velocity continuation that maps seismic images
associated with one background model to another virtually for free.
While being trained with only 200 background and seismic image pairs,
this surrogate is able to accurately predict seismic images associated
with new background models, thus accelerating seismic imaging
uncertainty quantification. We support our method with a realistic data
example in which we quantify seismic imaging uncertainties using a
Fourier neural operator surrogate, illustrating how variations in
background models affect the position of reflectors in a seismic image.
\end{abstract}

\vspace*{-0.4cm}

\section{Introduction}\label{introduction}

\vspace*{-0.25cm}

Seismic imaging involves estimating the short-wavelength component of
the Earth's subsurface squared-slowness model---known as the seismic
image---given shot records and an estimation of the smooth background
squared-slowness model. This linearized imaging problem is challenged by
the computationally expensive forward operator as well as presence of
measurements noise, linearization errors, modeling errors, and the
nontrivial nullspace of the linearized forward Born modeling operator
\citep{lambare1992iterative, schuster1993least, nemeth1999least}. These
challenges highlight the importance of uncertainty quantification (UQ)
in seismic imaging, where instead of finding one seismic image estimate,
a distribution of seismic images is obtained that explains the observed
data \citep{tarantola2005inverse}, consequently reducing the risk of
data overfit and enabling UQ
\citep{malinverno2006two, MartinMcMC2012, chevron2017, zhu2016bayesian, fang2018uqfip, stuart2019two, herrmann2019NIPSliwcuc, siahkoohi2020EAGEdlb, siahkoohi2020SEGuqi, siahkoohi2021Seglbe, siahkoohi2021deep}.

The seismic imaging uncertainty can be attributed to two main sources
\citep{thore2002structural, osypov2013model, Ely2018}: (1) errors in the
data, which include measurement and linearization errors; and (2)
modeling errors, which include errors in the estimation of the
background squared-slowness model. In this paper, we focus on
uncertainties with respect to the background model as it has the main
contributing factor to imaging uncertainty due to its effect on
reflector positioning
\citep{fomel2014structural, poliannikov2016effect, Ely2018}. To this
end, we assume there are no measurement or linearization errors and
consider the map from shot records to the seismic image to be
represented by the deterministic reverse-time migration (RTM) algorithm.
In this setup, quantifying the uncertainty in seismic imaging---due to
errors in the background model---involves computing numerous RTMs with
all the background model posterior samples
\citep{zhu2016bayesian, fang2018uqfip}. Due to the high-dimensionality
of seismic images, the number of seismic imaging posterior samples
required to obtain accurate estimations of the imaging posterior moments
is large, making UQ with respect to the background model computationally
expensive for 2D and unfeasible for large 3D problems. These costs can
be alleviated via velocity continuation methods
\citep{fomel2003time, fomel2003velocity, duchkov2009velocity, van2012wave, yang2021low}
that map seismic images associated with one background model to another,
without directly solving the imaging problem. Velocity continuation
methods are designed to be frugal compared to computing a new RTM for
each new background model, which makes them suitable for large-scale
seismic imaging uncertainty quantification
\citep{fomel2014structural, poliannikov2016effect}.

While several velocity continuation methods have been proposed
\citep{fomel2003time, fomel2003velocity, duchkov2009velocity, van2012wave, yang2021low},
they typically involve solving partial differential equations (PDEs),
which can be still costly in the context of seismic imaging UQ. To
address this challenge, we propose a neural network surrogate for
velocity continuation that is capable of mapping seismic images
associated with one background model to another with negligible
computational cost. Motivated by the success of Fourier neural operators
\citep[FNOs,][]{li2021fourier} in approximating the solution operator of
PDEs \citep{toledo2021deep, konuk2021physics, kovachki2021universal}, we
chose them as the architecture for our neural network surrogate. Due to
our main interest in accelerating velocity continuation in the context
of UQ, we train a survey-specific FNO that acts as a surrogate for
velocity continuation for the specific survey at hand. This choice,
while not offering generalizations across different survey areas in the
Earth, can speed up seismic imaging UQ for the survey at hand, which
involves computing seismic images associated with many background model
posterior samples. We show that the FNO can be trained using $200$ pairs
of background and seismic image pairs, making the survey-specific
training procedure computationally viable. To scale this method to
industry size problems, transfer learning
\citep{yosinski2014transferable} can further reduce the upfront costs of
training the FNO. After training, the FNO can be used to obtain samples
from the imaging posterior almost free of cost.

In the next section, we describe seismic imaging by introducing the
forward Born modeling operator, through which the seismic image relates
to the background model. Next, we define velocity continuation, followed
by describing our proposed FNO-based approach for accelerating seismic
imaging UQ. Finally, we evaluate the performance of the trained neural
operator on a realistic dataset, and we demonstrate how the imaging
uncertainty affects the positioning of the reflectors in the seismic
image.

\vspace*{-0.4cm}

\section{Theory}\label{theory}

\vspace*{-0.25cm}

We introduce a deep-network surrogate for velocity continuation in order
to enable faster quantification of uncertainty---due to errors in the
background velocity model---in seismic imaging. We begin with an
introduction to seismic imaging and the linearized forward model
associated with it.

\subsection{Seismic imaging}\label{seismic-imaging}

\vspace*{-0.15cm}

The inverse problem that we tackle involves the process of estimating
the short-wavelength component of the Earth's unknown subsurface
squared-slowness model given measurements recorded at the surface. This
problem, also known as seismic imaging, can be formulated as a linear
inverse problem by linearizing the nonlinear relationship between shot
records and the squared-slowness model, governed by the wave-equation.
In its simplest acoustic form, the linearization with respect to the
slowness model---around a background squared slowness model
$\B{m}_0$---leads to a linear inverse problem for estimating the ground
truth seismic image $\delta \B{m}^{\ast}$ with the following forward
model,
\begin{equation}
\delta \B{d}_i = \B{J}(\B{m}_0, \B{q}_i)
    \delta \B{m}^{\ast}.
\label{linear-fwd-op}
\end{equation}
 In the above expression,
$\delta \B{d}=\left\{\delta \B{d}_{i}\right\}_{i=1}^{n_s}$ are $n_s$
linearized shot records and $\B{J}(\B{m}_0, \B{q}_i)$ represents the
linearized Born scattering operator. This operator is parameterized by
the source signature $\B{q}_{i}$ and the background squared-slowness
model $\B{m}_0$, which is typically estimated in the previous inversion
steps \citep{zhu2016bayesian, fang2018uqfip}. In this work, we focus on
potential inaccuracies in the background model, the main source of
variability and uncertainty in seismic imaging
\citep{fomel2014structural, poliannikov2016effect, Ely2018}. Therefore,
we assume there are not measurement and linearization errors, which
leads to a deterministic mapping from
$\left\{\delta \B{d}_{i}\right\}_{i=1}^{n_s}$ to $\delta \B{m}$ for a
given background model. We use RTM for this deterministic map, which
involves applying the adjoint linearized Born scattering operator to the
linearized shot records for all source experiments,
\begin{equation}
\delta \B{m}_{\text{RTM}} = \sum_{i=1}^{n_s} \B{J}(\B{m}_0,
  \B{q}_i)^{\top} \delta\B{d}_i.
\label{rtm}
\end{equation}
 Since this map is deterministic, the uncertainty in the background
model can be translated into uncertainty in seismic imaging by
evaluating Equation~\ref{rtm} for the background models sampled from the
posterior distribution $p(\B{m}_0\mid \B{d})$, where $\B{d}$ represents
the shot records with no linearization (the field data)
\citep{zhu2016bayesian, fang2018uqfip}. Bayesian inference involving
high-dimensional posterior distributions, for example seismic imaging
and full-waveform inversion, requires many samples from the posterior
distribution for accurate Monte Carlo approximation of high-dimensional
integrals \citep{gelman_rubin_92}. As a result, the mapping in
Equation~\ref{rtm} must be evaluated over numerous background model
samples from $p(\B{m}_0\mid \B{d})$, which is computationally expensive
due to costs of applying
$\B{J}(\B{m}_0,\B{q}_i)^{\top},\,i=1,\ldots,n_s$. To address this
computational challenge, our proposed method involves a FNO-based
velocity continuation approach that is capable of mapping seismic images
associated with one background model to another, effectively replacing a
costly demingration-migration. We describe this approach in the next
section.

\subsection{Velocity continuation}\label{velocity-continuation}

\vspace*{-0.15cm}

At its core, velocity continuation is a process which alters a seismic
image based on changes in the background model
\citep{fomel2003time, fomel2003velocity, van2012wave, yang2021low}.
Using this technique, an initial seismic image associated with an
initial background model $\B{m}_{\text{init}}$ is altered to approximate
the seismic image associated with a target background model
$\B{m}_{\text{target}}$ without computing Equation~\ref{rtm}. This
process can be interpreted as a map \citep{duchkov2009velocity}:
\begin{equation}
\mathcal{T}_{\left(\B{m}_{\text{init}}, \B{m}_{\text{target}}\right)}:
\delta \mathcal{M} \to \delta \mathcal{M},
\label{velocity-continuation}
\end{equation}
 where $\delta \mathcal{M}$ denotes the space of seismic images, and the
velocity continuation map is parameterized by $\B{m}_{\text{init}}$ and
$\B{m}_{\text{target}}$. We take advantage of recent advances in deep
learning to train a surrogate model for velocity continuation that
approximates
$\mathcal{T}_{\left(\B{m}_{\text{init}},\B{m}_{\text{target}}\right)} = \B{J}(\B{m}_{\text{target}},\B{q}_i)^{\top}(\B{J}(\B{m}_{\text{init}}, \B{q}_i)^{\top})^{\dagger}$,
which can be evaluated virtually for free instead of solving four PDEs.
Through this approach, we are able to significantly accelerate velocity
continuation, trading wave-equation solves for a simple neural network
inference. This is of considerable importance in the context of seismic
imaging UQ. Before demonstrating the benefits of our approach, we
introduce FNOs in the context of velocity continuation.

\subsection{Fourier neural operators for velocity
continuation}\label{fourier-neural-operators-for-velocity-continuation}

\vspace*{-0.15cm}

In light of their success in learning mesh-free solution operators to
PDEs \citep{toledo2021deep, konuk2021physics, kovachki2021universal}, we
choose FNOs \citep{li2021fourier} as a surrogate model for velocity
continuation---a process that can be interpreted as a double linearized
PDE solve. The main components of FNOs are the Fourier layers, which
involve a Fourier transform over the spatial dimensions of their input,
followed by a learned pointwise multiplication and an inverse Fourier
transform. These layers act as long-kernel convolutional layer, akin to
pseudo-spectral methods, which explains the representation power of FNOs
\citep{kovachki2021universal}. To train a FNO as a surrogate model for
velocity continuation, we define it as a map
\begin{equation}
\mathcal{G}_{\B{w}} : \mathcal{M} \times \delta \mathcal{M} \to \delta \mathcal{M},
\label{fno-map}
\end{equation}
 where $\mathcal{G}_{\B{w}}$ denotes the FNO with weights $\B{w}$, and
$\mathcal{M}$ is the space of background models. In words, we design
$\mathcal{G}_{\B{w}}$ as a neural network that takes as input the target
background model and the initial seismic image, and outputs the target
seismic image. This choice is analogous to the structure of the velocity
continuation map
$\mathcal{T}_{\left(\B{m}_{\text{init}},\B{m}_{\text{target}}\right)}$
(cf.~Equation~\ref{velocity-continuation}), except that we fix the
initial background model $\B{m}_{\text{init}}$ to an arbitrary
background model posterior sample, hence making the dependence of
$\mathcal{G}_{\B{w}}$ to $\B{m}_{\text{init}}$ implicit. With this
choice of inputs and output for $\mathcal{G}_{\B{w}}$, the FNO's task is
to perturb the input initial seismic image according to the provided
target background model in order to predict the target seismic image.

Due to our interest in accelerating velocity continuation in the context
of UQ, we train a survey-specific FNO that acts as a surrogate for
velocity continuation for the specific survey at hand. This choice,
while not offering generalizations across different survey areas in the
Earth, can speed up seismic imaging UQ for the survey at hand, which
involves computing seismic images associated with many posterior
background model samples $\B{m}_0 \sim p(\B{m}_0\mid \B{d})$. By
training the FNO on a small number of these background model and seismic
image pairs, i.e., approximately $200$, we can accelerate the velocity
continuation process for the rest of the background model posterior
samples while limiting the risk of introducing generalization errors due
to the strong heterogeneity of Earth. To achieve this, we construct a
set of $N$ training input-output pairs in the form of
\begin{equation}
\bigg\{ \Big ( \big(\B{m}_{0}^{(i)},\delta
\B{m}_{\text{init}} \big),\delta \B{m}_{\text{RTM}}^{(i)} \Big) \ \Big | \
i=1,\ldots,N \bigg \},
\label{fno-training-pairs}
\end{equation}
 where $(\B{m}_{0}^{(i)},\delta \B{m}_{\text{init}})$ is the input
target background and initial seismic image training pair, and
$\delta \B{m}_{\text{RTM}}^{(i)}$ is the associated target seismic
image. Training involves minimizing the squared $\ell_2$-norm of the
difference between the FNO output and the target seismic image with
respect to FNO weights,
\begin{equation}
\B{w}^{\ast}  = \argmin_{\B{w}}\, \frac{1}{N} \sum_{i=1}
\| \mathcal{G}_{\B{w}}(\B{m}_{0}^{(i)}, \delta \B{m}_{\text{init}}) - \delta
\B{m}_{\text{RTM}}^{(i)} \|_2^2.
\label{fno-train}
\end{equation}
 We solve optimization problem~\ref{fno-train} with the Adam stochastic
optimization algorithm \citep{kingma2014adam}. After training, the
trained FNO approximates the velocity continuation map for a fixed
homogeneous initial background model, i.e.,
\begin{equation}
\mathcal{G}_{\B{w}^{\ast}}(\B{m}_{\text{target}},
\delta \B{m}_{\text{init}}) \approx
\mathcal{T}_{\left(\B{m}_{\text{init}}, \B{m}_{\text{target}}\right)}
(\delta \B{m}_{\text{init}}) = \delta \B{m}_{\text{target}},
\label{fno-approx}
\end{equation}
 where $\B{m}_{\text{init}}$ is fixed to an arbitrary background model
posterior sample and $\delta \B{m}_{\text{target}}$ denotes the seismic
image associated with $\delta \B{m}_{\text{init}}$. Given background
model samples $\B{m}_0 \sim p(\B{m}_0\mid \B{d})$ as input, the FNO
outputs samples from the imaging posterior,
$p(\B{m}_0\mid \delta \B{d})$. This process accelerates seismic imaging
uncertainty quantification as no further RTMs (Equation~\ref{rtm}) need
to be computed. Training the FNO with as little as $200$ training pairs
makes the survey-specific training procedure computationally viable. To
further accelerate the process, training of the FNO can be started as
the same time as the background model posterior sampling phase, using
the already collected posterior samples as training data. In the next
section, we show the results of approximating the velocity continuation
map with a FNO via a quasi-real seismic experiment.

\vspace*{-0.4cm}

\section{Numerical experiments}\label{numerical-experiments}

\vspace*{-0.25cm}

The purpose of the presented numerical experiments here is to
demonstrate the ability to approximate the velocity continuation map
(Equation~\ref{velocity-continuation}) using a FNO. We further show how
this trained FNO can be used for UQ by showing the effect of imaging
uncertainty on the positioning of the reflectors. We begin by describing
the training setup, including the seismic acquisition geometry.

\subsection{Acquisition geometry and training
configuration}\label{acquisition-geometry-and-training-configuration}

\vspace*{-0.15cm}

\begin{figure}
\centering
\subfloat[\label{m0-var}]{\includegraphics[width=0.500\hsize]{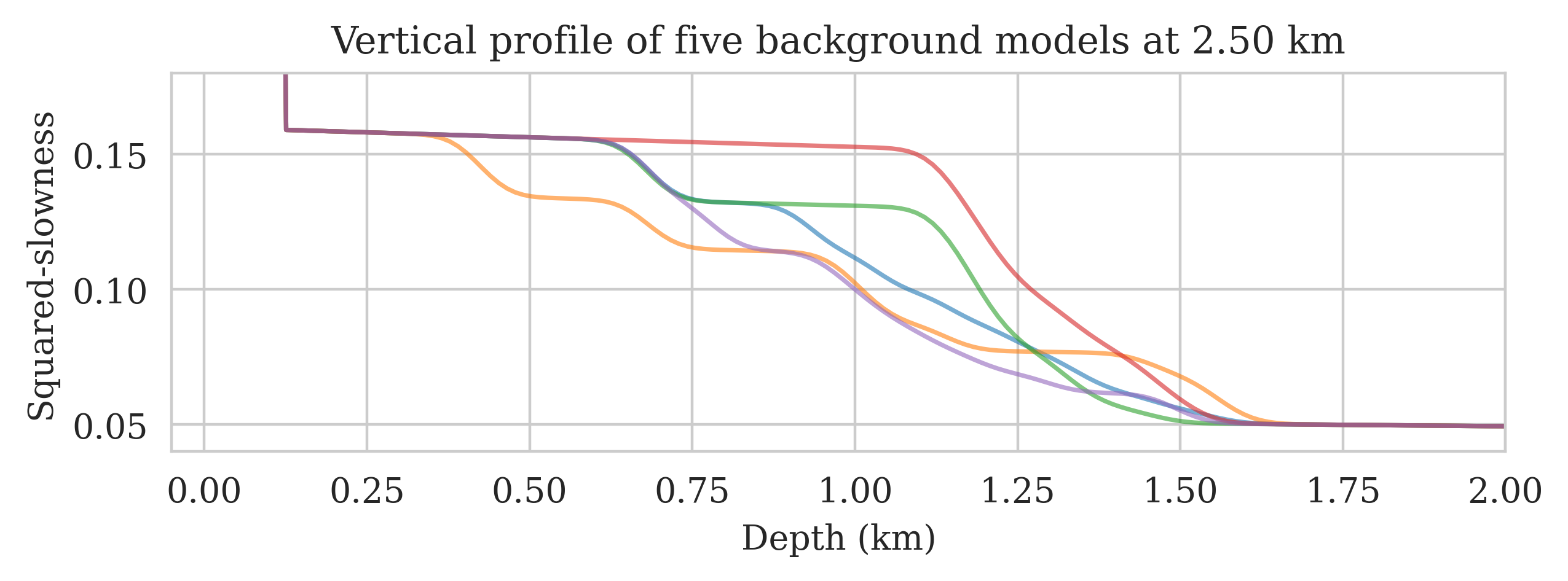}}
\subfloat[\label{dm-var}]{\includegraphics[width=0.500\hsize]{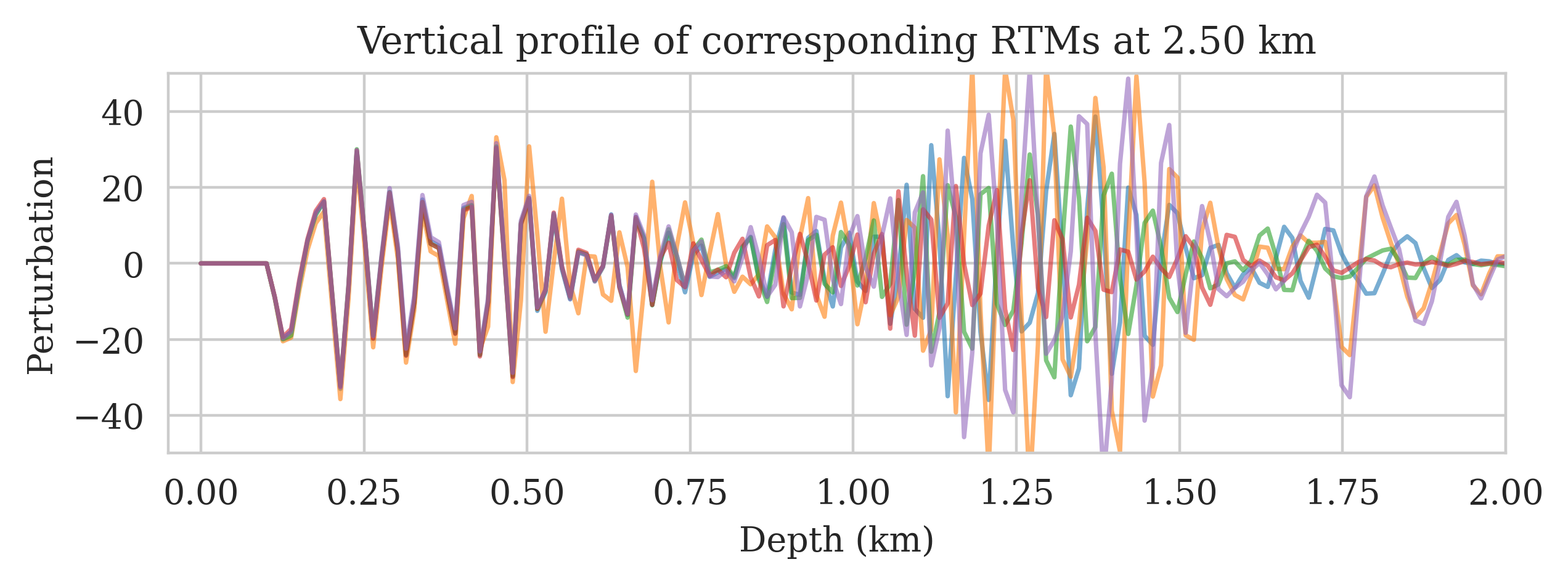}}
\caption{Variation among five (a) background; and (b) seismic images
(RTMs), plotted as a vertical profile at
$2.50\,\mathrm{km}$.}\label{variation}
\end{figure}

Our examples involve imaging a 2D subset of the
\href{https://wiki.seg.org/wiki/Parihaka-3D}{Parihaka}
\citep{Veritas2005, WesternGeco2012} prestrack Kirchhoff migration field
dataset. We use this 2D section to create linearized data according to
the linear forward model in Equation~\ref{linear-fwd-op}, where we
consider no measurement and linearization errors. The model is
discretized on a $12.5\,\mathrm{m}$ vertical and $20\,\mathrm{m}$
horizontal grid, and the data is acquired with $102$ equally spaced
sources and $204$ fixed receivers. We use a Ricker wavelet with a
central frequency of $30\, \mathrm{Hz}$ as the source signature. For
presentation purposes, we augment a $125\, \mathrm{m}$ water column on
top of these models to limit the near source imaging artifacts. We
create background models by assuming access to an oracle, which provides
geologically consistent background models with the 2D section. Given
$200$ background models, we migrate the simulated seismic data via
Equation~\ref{rtm} to obtain the corresponding seismic images.
Figures~\ref{m0-var} and~\ref{dm-var} show the vertical profile of five
randomly selected background and seismic images at $2.50\,\mathrm{km}$,
respectively. These images, displaying strong amplitude and phase
differences, highlight the importance of quantifying the uncertainty in
imaging when dealing with errors in the background model. To reduce the
costs associated with UQ, we train the FNO surrogate using the $200$
training pairs (cf.~Equation~\ref{fno-training-pairs}) for $500$ epochs
with the Adam optimizer \citep{kingma2014adam}.

\subsection{Results}\label{results}

\vspace*{-0.15cm}

\begin{figure}
\centering
\subfloat[\label{m0}]{\includegraphics[width=0.500\hsize]{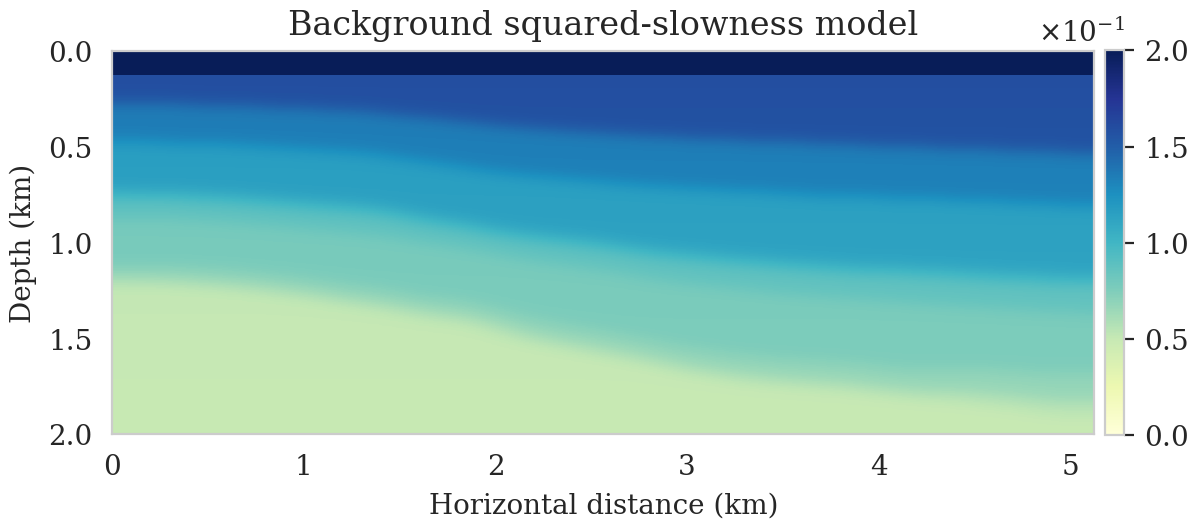}}
\subfloat[\label{desired}]{\includegraphics[width=0.500\hsize]{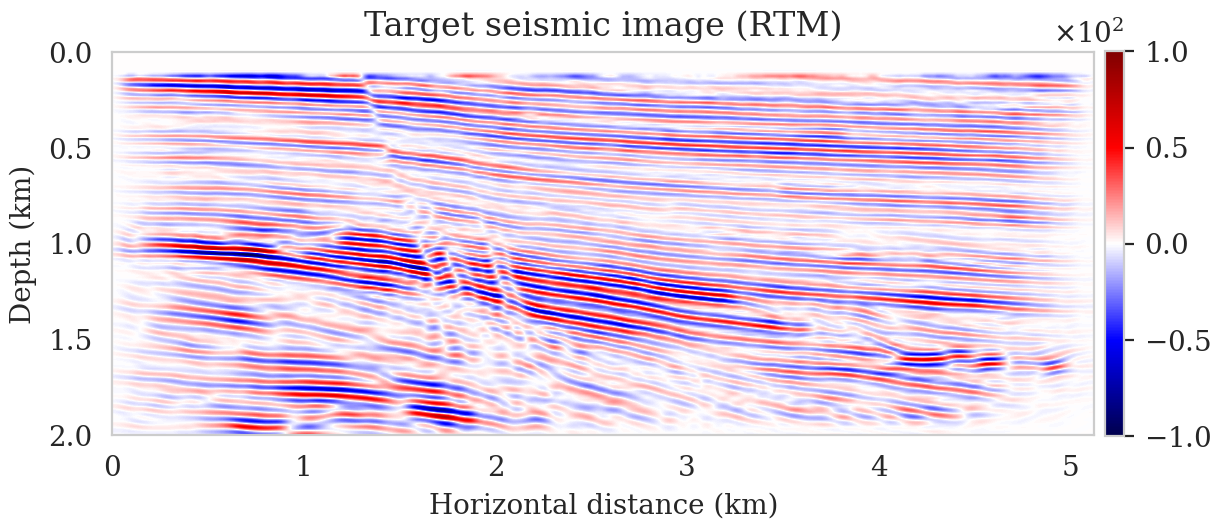}}
\\
\subfloat[\label{predicted}]{\includegraphics[width=0.500\hsize]{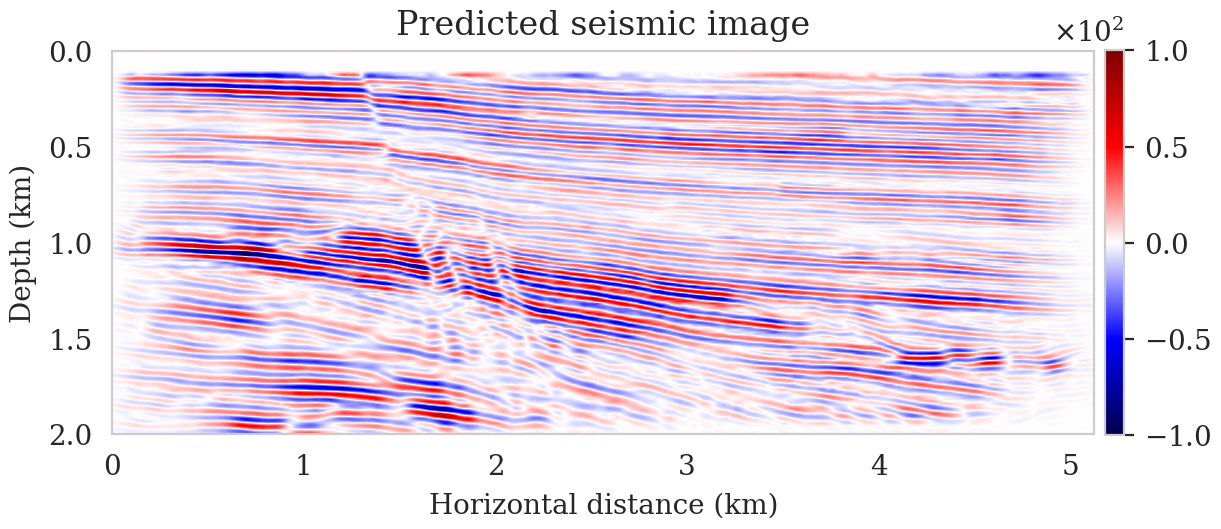}}
\subfloat[\label{error}]{\includegraphics[width=0.500\hsize]{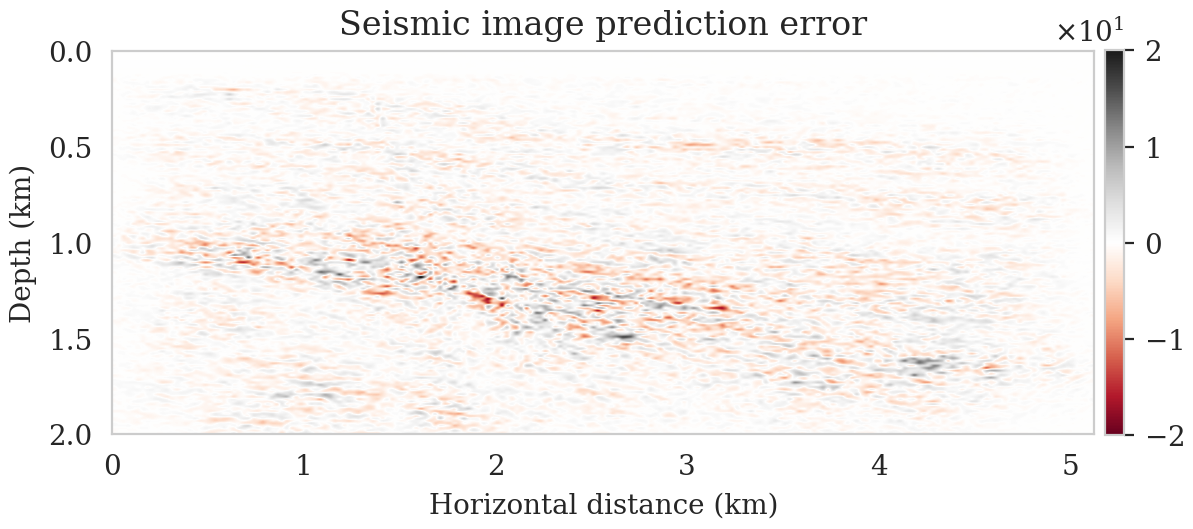}}
\\
\subfloat[\label{trace-1}]{\includegraphics[width=0.500\hsize]{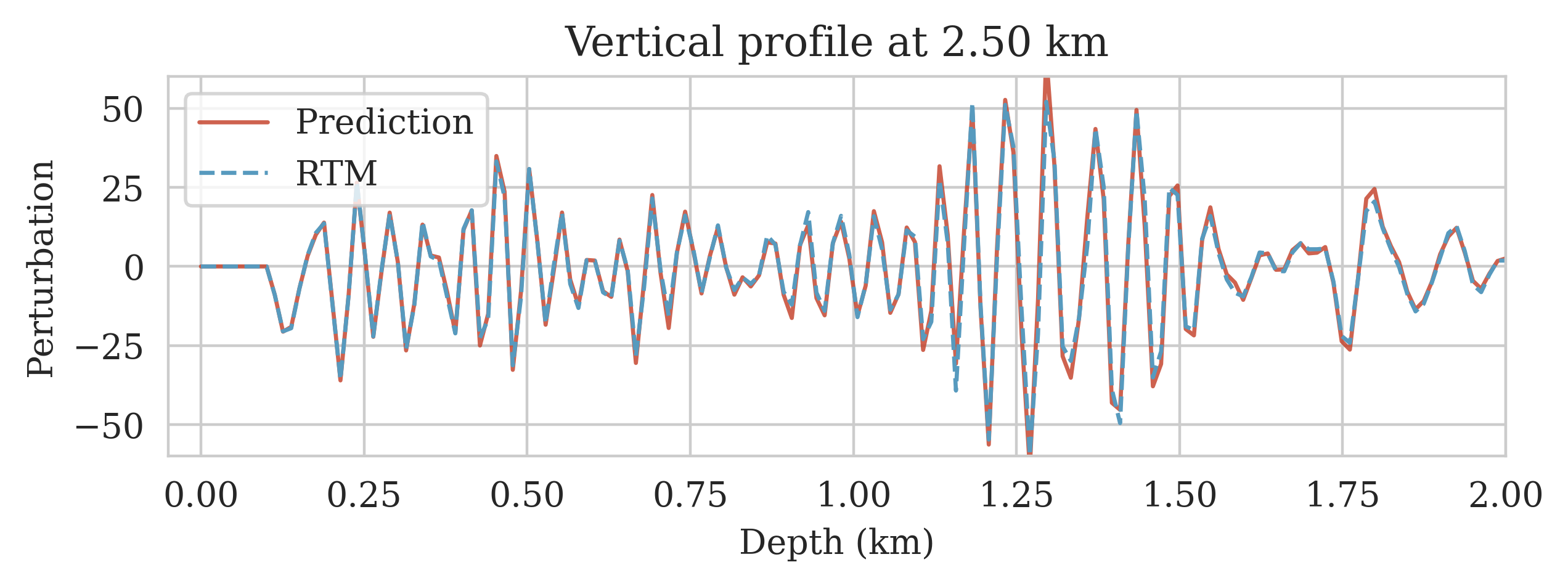}}
\subfloat[\label{trace-2}]{\includegraphics[width=0.500\hsize]{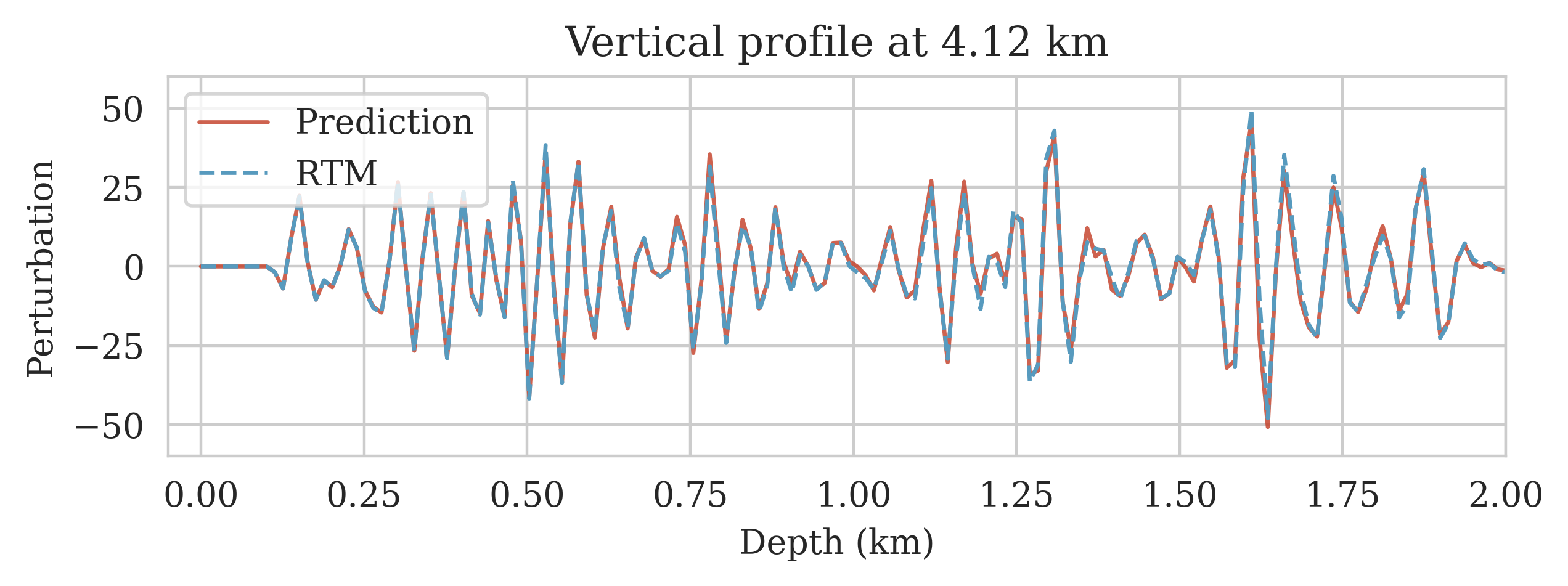}}
\caption{Velocity continuation with FNOs. Target (a) background; and (b)
seismic images. (c) Predicted seismic image with the FNO. (d) Difference
between target and prediction. (e) and (f) Vertical profile comparisons
between target (dashed blue) and predicted (red) seismic
images.}\label{example-1}
\end{figure}

\begin{figure}
\centering
\includegraphics[width=0.750\hsize]{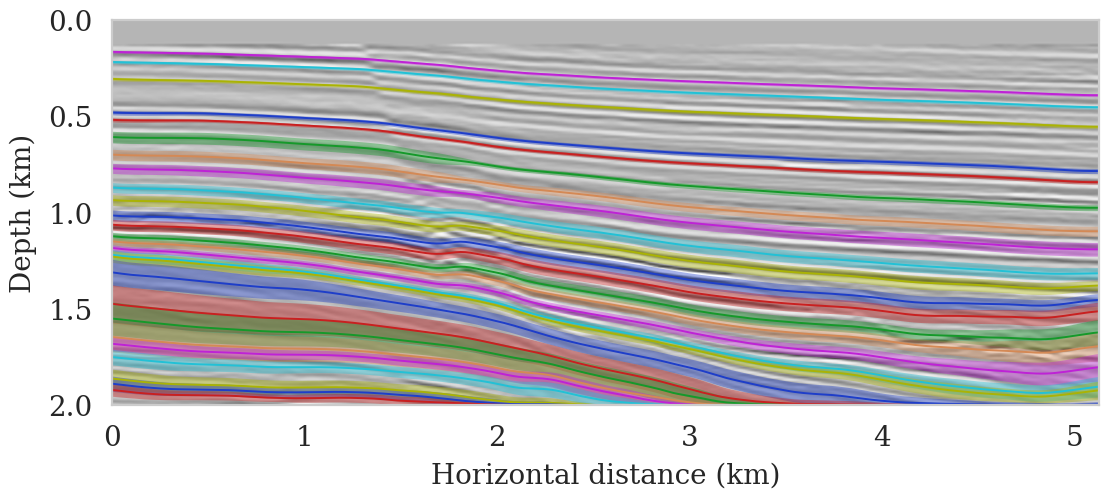}
\caption{Uncertainty in the tracked horizons.}\label{example-2}
\end{figure}

To evaluate the accuracy of the trained FNO in predicting seismic
images, we compare its output to the target seismic image, i.e., the RTM
image obtained via Equation~\ref{rtm} when using the target background
model (Figure~\ref{m0}) to parameterize the Born scattering forward
operator. This is conducted over the testing dataset, which is derived
using the same procedure as the training dataset using the background
model creating oracle. Figures~\ref{desired} and~\ref{predicted} show
the target and predicted seismic images, respectively, and
Figures~\ref{error} includes the difference between them. We observe
that the network has accurately predicted the target image, where errors
are mostly due to amplitude differences. The accuracy of the prediction
in terms of phase and amplitude can be further confirmed by focusing on
two vertical profiles, at $2.50\,\mathrm{km}$ (Figure~\ref{trace-1}) and
$4.125\,\mathrm{km}$ (Figure~\ref{trace-2}) horizontal locations, which
include regions with fault and torturous reflectors. We use the trained
FNO for UQ by providing testing background models and predicting the
associated seismic images. We visualize the obtained uncertainties by
showing the variability in the location of reflectors, determined via an
automatic horizon tracking software \citep{wu2018least}. We pass the
seismic images predicted by the FNO to the horizon tracker for $25$
selected horizons. As a result, we obtain multiple instances of each
horizons, from which we compute pointwise mean and standard deviations.
Figure~\ref{example-2} indicates the result where the solid lines
correspond to the mean among different instances of each horizons and
shaded areas indicate the mean plus and minus the pointwise standard
deviation. These shaded areas indicate uncertainties in the location of
the reflectors, which are due to the variability in the background
models. As expected, we find a general increase of uncertainty with
depth. We also observe that the areas of high uncertainty are correlated
with areas of poor illumination, faults and tortuous reflectors.

In this example, we use Devito \citep{devito-compiler, devito-api} for
the wave-equation based simulations. We based our PyTorch FNO
implementation on the
\href{https://github.com/zongyi-li/fourier_neural_operator}{original
implementation}. The code to reproduce our results are made available on
\href{https://github.com/slimgroup/velocity_continuation_with_FNOs}{GitHub}.

\vspace*{-0.4cm}

\section{Conclusions and discussion}\label{conclusions-and-discussion}

\vspace*{-0.25cm}

Quantifying the uncertainty in seismic imaging due to errors in the
background model involves solving many seismic imaging problems that
vary in the parameterization of the background model of the forward
operator. To reduce the computational cost of this process---mainly due
to the computational costs of the forward operator---we proposed to
train a survey-specific Fourier neural operator surrogate that mimics
velocity continuation. This surrogate model maps seismic images
associated with one background model to another virtually for free,
which has the benefit of accelerating uncertainty quantification. We
showed that this surrogate model can be trained with as few as $200$
training pairs while still providing a good seismic image prediction
accuracy. Further research is required in training a reliable global
surrogate, being able to generalize across other survey areas and more
realistic physics.

\vspace*{-0.4cm}

\section{Acknowledgment}\label{acknowledgment}

\vspace*{-0.25cm}

This research was carried out with the support of Georgia Research
Alliance and partners of the ML4Seismic Center.

\bibliography{abstract}

\end{document}